# Magnetic phase diagram of $Cr_2Te_3$ revisited by ac magnetostrictive coefficient

Long Zhang,[1] Zhongzhu Jiang,[2] Yugang Zhang,[1] Jing Zhang,[1] Aifeng Wang,[1] Mingquan He,[1] Y. P. Sun,[2,3,4] Xuan Luo,[2,a)] and Yisheng Chai[1,a)]

[1]*Low Temperature Physics Laboratory, College of Physics, Chongqing University, Chongqing 401331, China*

[2]*Key Laboratory of Materials Physics, Institute of Solid State Physics, HFIPS, Chinese Academy of Sciences, Hefei 230031, China.*

[3]*Anhui Province Key Laboratory of Low-Energy Quantum Materials and Devices, High Magnetic Field Laboratory, HFIPS, Chinese Academy of Sciences, Hefei 230031, China.*

[4]*Collaborative Innovation Center of Advanced Microstructures, Nanjing University, Nanjing 210093, China.*

[a)]Authors to whom correspondence should be addressed: xluo@issp.ac.cn and yschai@cqu.edu.cn,

**Abstract:**

Two-dimensional (2D) magnetic materials have attracted considerable interest owing to their potential applications in spintronics and fundamental investigations into low-dimensional magnetism. $Cr_2Te_3$, a quasi-2D non-van der Waals magnet, exhibits a complex magnetic phase diagram due to competing magnetic interactions within and between layers. However, the precise nature and evolution of these magnetic phases remain unclear. Here, we utilize an ultrahigh-sensitive composite magnetoelectric technique, which probes the ac magnetostrictive coefficient, to systematically explore the temperature–magnetic field phase diagram of $Cr_2Te_3$ single crystals. Our results reveal the coexistence of multiple magnetic phases, including canted ferromagnetic, antiferromagnetic, and paramagnetic states. Another canted ferromagnetic phase and a possible triple point have been proposed. The updated phase diagram provides deeper insights into the specific spin configurations associated with each phase. These findings also highlight the decoupled magnetic ordering between the Cr1/Cr3 layers and the Cr2 layer near the magnetic ordering temperature.

Two-dimensional (2D) magnetic materials have emerged as a vibrant research area in condensed matter physics, offering a versatile platform for exploring magnetic exchange interactions and presenting transformative potential for spintronic applications.[1-3] While van der Waals ferromagnets such as $Cr_2Ge_2Te_6$ and $CrI_3$ have demonstrated notable device-oriented functionalities,[4,5] 2D antiferromagnetic (AFM) materials also exhibit unique promise, particularly as spin-polarized sources or filters.[6] Integrating antiferromagnets into heterostructures or superlattices enables selective modulation of spin-polarized currents, facilitating advanced spintronic device designs.[7] However, a critical limitation remains: most 2D magnetic systems, including ferromagnetic (FM) semiconductors (e.g., $Cr_2Ge_2Te_6$, $CrI_3$) and metallic ferromagnets (e.g., $Fe_3GeTe_2$), experience significantly suppressed magnetic ordering temperatures at reduced thicknesses, restricting their practical utility.[8] Remarkably, $Cr_2Te_3$—a non-van der Waals material exhibiting quasi-2D structural characteristics—defies this common limitation. Yao *et al.* recently reported that reducing $Cr_2Te_3$ thickness to 7.1 nm raises its FM Curie temperature to 280 K, close to room temperature, representing a significant step toward spintronic applications.[9] This thickness-dependent enhancement starkly contrasts with typical 2D magnets, making $Cr_2Te_3$ an exceptional candidate for both fundamental studies and device integration.

Bulk $Cr_2Te_3$ crystallizes in a trigonal structure with space group P-3$1c$ (No. 163), featuring a layered arrangement of Cr and Te atoms, as illustrated in Fig. 1(a). Structurally, it comprises alternating $CrTe_2$ (Cr1 and Cr3) layers interleaved with Cr-deficient (Cr2) layers. Magnetically, bulk $Cr_2Te_3$ exhibits complex behavior, hosting competing FM and AFM phases.[10] Neutron scattering measurements indicate magnetic moments of −0.14 $\mu_B$, 2.78 $\mu_B$, and 2.53 $\mu_B$ for Cr2, Cr1, and Cr3 sites at 85 K, respectively, suggesting a canted FM ground state.[11] However, through comprehensive studies combining specific heat, variable-temperature X-ray diffraction, thermal expansion measurements, and theoretical calculations, Luo et al. identified a canted FM (CFM) ground state.[10] Upon warming, $Cr_2Te_3$ transitions into an AFM phase at approximately 160 K through a first-order phase transition. Under moderate magnetic fields (*H*) along the *c*-axis, this transition temperature (*T*) increases to 178 K, accompanied by a crossover to second-order behavior. A subsequent second-order AFM-to-paramagnetic (PM) transition occurs at ~181 K, with the AFM phase entirely suppressed at a critical field of 0.48 T. However, the detailed spin configuration within the AFM phase remains unresolved, and the critical phenomena around the first-to-second-order crossover, including fluctuations near the tricritical point, remain largely unexplored. Clarifying these aspects could provide new avenues to engineer magnetic phase transitions in low-dimensional materials, further establishing $Cr_2Te_3$ as an attractive candidate for fundamental research and practical applications.

Recently, we developed an ultrahigh-sensitive composite magnetoelectric (ME) technique capable of probing intricate magnetic phase transitions in diverse systems, including 2D magnetic materials,[12] magnetic skyrmions,[13,14] and Kitaev quantum spin liquid candidates.[15] Its high sensitivity mainly due to the fact that the ME signal is directly proportional to the ac magnetostrictive coefficient of the sample.[16] Therefore, this technique has been applied to: (1) distinguish between first- and second-order phase transitions via its imaginary component,[12] (2) identify triple and tricritical points associated with magnetic phase transitions,[12,15] and (3) reveal domain wall motion or rotation mechanisms during magnetization processes.[12] Thus, it is exceptionally well-suited for revisiting and clarifying the complex magnetic phase diagram of $Cr_2Te_3$.

In this letter, we revisit the *H-T* phase diagram of $Cr_2Te_3$ by the aforementioned composite ME technique. The measurements reveal previously unresolved phase coexistence phenomena, specifically, the overlapping CFM and AFM phases, as well as coexistence of AFM and PM phase in the higher temperature region. Notably, we find another CFM2 phase and propose a triple point within the AFM phase. The above findings suggest that the Cr1/Cr3 layers and Cr2 layer have decoupled magnetic sublattice ordering in this quasi-2D system. These findings significantly advance the understanding of complex magnetic interactions in $Cr_2Te_3$. The updated phase diagram provides essential insights for manipulating magnetic states in this material system through thermal and field control parameters.

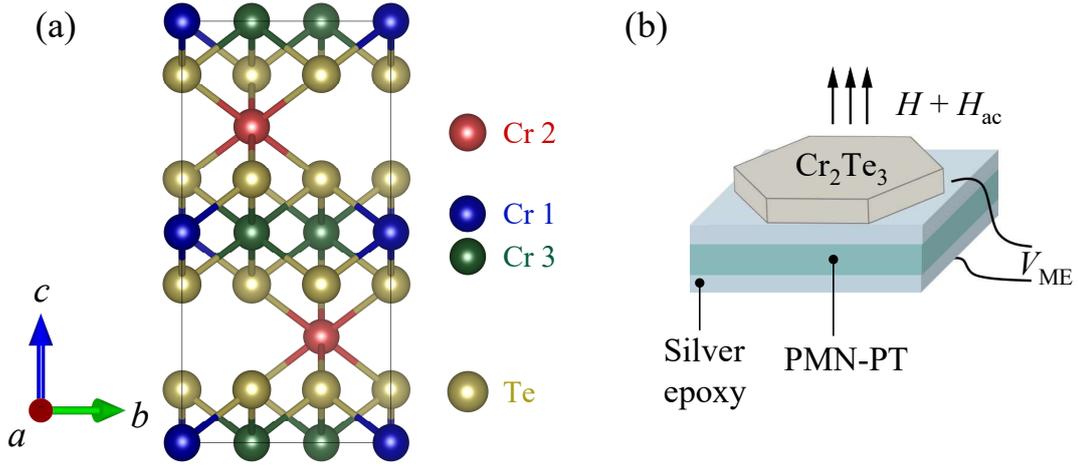

FIG. 1. (a) The crystal structure of $Cr_2Te_3$. (b) The configuration of composite ME technique of $Cr_2Te_3$/PMN-PT. The ac magnetic field $H_{ac}$ is applied parallel to the dc field $H$ and parallel to the $c$-axis of $Cr_2Te_3$.

The $Cr_2Te_3$ single crystal samples were grown using the self-flux method described in Ref.10. The crystal structure, magnetization, resistivity, specific heat, and thermal expansion of the studied samples have already been characterized in our previous study.[10] For the composite ME technique, the $Cr_2Te_3$ single crystal is mechanically bonded with a [001]-cut $0.7Pb(Mg_{1/3}Nb_{2/3})O_3–0.3PbTiO_3$ (PMN-PT) single crystal by silver epoxy (Epo-Tek H20E, Epoxy Technology, Inc.), as shown in Fig. 1(b). In this configuration, the magnetostriction $\lambda$ of $Cr_2Te_3$ will be converted into electrical signal across the PMN-PT. To enhance the sensitivity, a small ac magnetic field $H_{ac} = 1$ Oe is applied as a driven field and the excited ac electric voltage $V_{ME}$ on the PMN-PT is measured by a lock-in amplifier with a commercial sample insert (Multifield Corp.). Here, $V_{ME}$ is a direct indicator of the magnetostrictive coefficient that:[15]

$$V_{ME} \propto k \frac{dE}{d\lambda}\frac{d\lambda}{dH} \propto \left(\frac{d\lambda}{dH}\right)_{ac} = \frac{d\lambda'}{dH} + i\frac{d\lambda''}{dH} \quad (1)$$

where $0 < k < 1$ reflecting the efficiency of strain transfer between the sample and PMN-PT, $dE/d\lambda$ being the piezoelectric coefficient of PMN-PT, $d\lambda'/dH$ and $d\lambda''/dH$ representing the real and imaginary part of $(d\lambda/dH)_{ac}$, and being used to represent the real and imaginary part of ME signals, respectively.

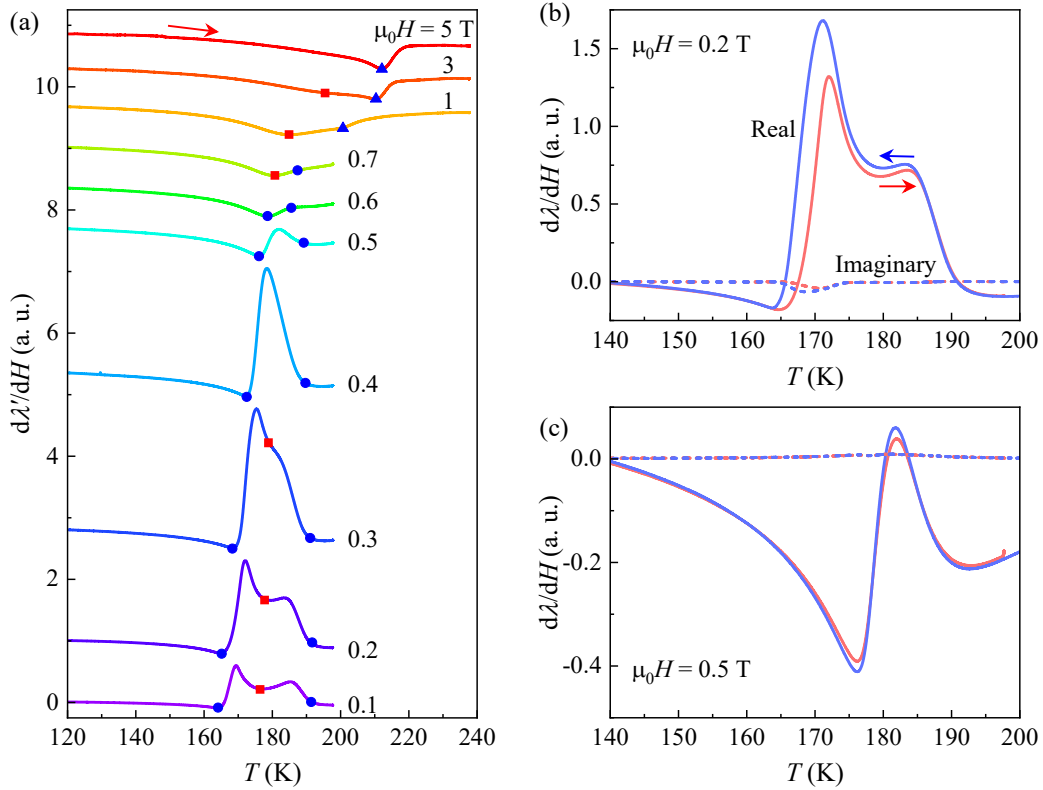

FIG. 2. (a) The $T$-dependent d$\lambda'$/d$H$ with the selected magnetic fields. Curves have been shifted vertically for clarity. The AFM and FM transition boundary are distinguished and marked with blue and red points, respectively. (b) and (c) display the warming and cooling processes of d$\lambda$/d$H$ under dc magnetic fields of 0.2 and 0.5 T, respectively. The solid (dotted) lines represent d$\lambda'$/d$H$ (d$\lambda''$/d$H$).

Figure 2(a) shows the temperature dependence of d$\lambda'$/d$H$ curves under selected external magnetic fields applied along the $c$-axis. For $\mu_0H < 0.4$ T, d$\lambda'$/d$H$ exhibits a characteristic of superimposed peak (positive) and valley (negative) features in the reported AFM region in between 165-190 K. Below 165 K, the d$\lambda'$/d$H$ gradually increases to zero with decreasing temperature ($T$), indicating a canted FM phase. For $\mu_0H = 0.2$ T data, d$\lambda'$/d$H$ shows clear thermal hysteresis around 165 K, pointing to a first order transition in nature between FM and AFM phases,[12] as shown in Fig. 2(b). In contrast, around 190 K near the falling edge of the curve, there is no hysteresis between cooling and warming data, implying a second order phase transition between AFM and PM phases. Meanwhile, the imaginary part d$\lambda''$/d$H$ shows distinct signals with negative peaks only near the FM-AFM phase transition boundary, reflecting dissipations due to a first-order transition.[12] For $0.4 \leq \mu_0H \leq 0.7$ T, the peak-valley feature in the AFM phase gradually becomes a single positive peak superimposed on a broad negative valley over a broad temperature region (120-200 K). In particular at $\mu_0 H$ = 0.5 T, the thermal hysteresis is absent in $d\lambda'/dH$ and $d\lambda''/dH$ goes to zero, as seen in Fig. 2(c). This indicates that the phase transition between FM and AFM changes from the first-order to the second-order above 0.4 T and should lead to a tricritical point at around 0.4 T and 172 K, in accordance with previous study.[15] With further increased magnetic field, the AFM peak finally disappears and merges to a single point at 0.7 T. For $\mu_0 H$ > 0.7 T, an FM-to-PM transition (marked by red solid square) emerges, characterized by a downward peak. This peak shifts to higher temperature with increasing field, with its intensity decreasing and its width broadening. Along with the gradual disappearance of this peak, another downward peak shows up from $\mu_0 H$ = 1 T at higher temperature. This new peak also goes to higher temperature with increasing field.

By comparing with data from other methods,[10] the positive peak in the $d\lambda'/dH$ between 165-190 K must come from the AFM magnetic ordering. The negative valley either deep in the AFM phase (< 0.4 T) or in high field data (>0.5 T) indicates an FM magnetic ordering. In this regard, the FM and AFM must coexist in some temperature and field regions. In the previous $T$-dependent magnetization measurements,[10] the slow rise of the magnetization with decreasing temperature in the AFM phase is consistent with this expectation. Such phase coexistence may also lead to the second high temperature transition above 1 T. We will discuss these points in the following text.

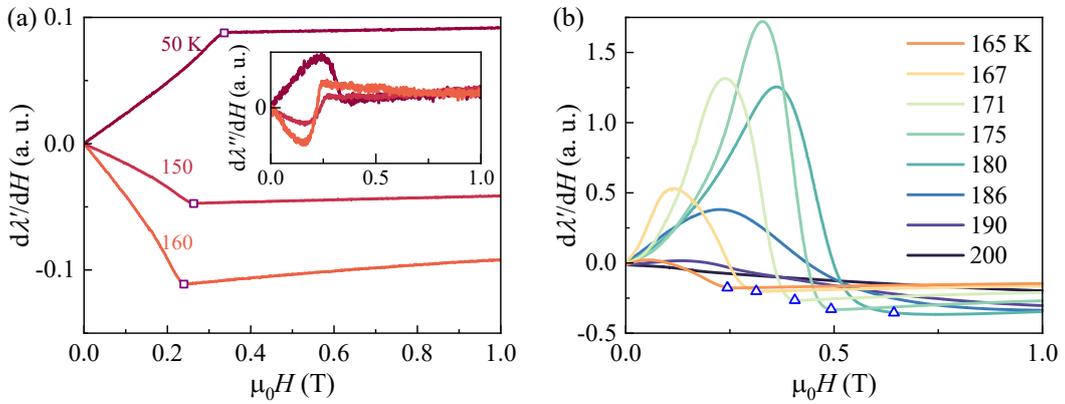

FIG. 3. The field-dependent $d\lambda'/dH$ at selected temperatures in the range of (a) 50-160 K and (b) 165-200 K, respectively. Two sets of transition fields are marked with hollow squares and triangles, respectively.

Figure 3 shows the field dependent d$\lambda'$/d$H$ at selected temperatures. Slightly below the FM-AFM transition temperature (<165 K), d$\lambda'$/d$H$ increases monotonically with the field, and saturated at about 0.25 T sharply. Compared with the published magnetization data,[10] such saturation behavior corresponds to a transition from a multi-domain canted FM state to a single-domain canted FM state (CFM). The magnetic moment of $Cr^{3+}$ (~2.04 $\mu_B$) derived from saturation magnetization is smaller than the theoretical value (~3.87 $\mu_B$),[10] which also support such CFM phase. The d$\lambda''$/d$H$ at those temperatures show clear peak features before the saturation field (Fig. 3(a), indicating the domain-wall motion induced dissipation in this field range. This interpretation is also corroborated with the similar imaginary peaks in $Cr_2Si_2Te_6$ at low fields.[12] As temperature decreases to 50 K, the saturation field increases while the magnitude of d$\lambda'$/d$H$ changes from negative to positive, even after saturation. Those sign changes are also reflected in the $T$ dependent data in Fig. 2. It is probably due to a temperature-induced modulation of the spin canting angle in the CFM state.

Upon entering the AFM state (>160 K, as shown in Fig. 3(b)), a positive peak in d$\lambda'$/d$H$ rises at low field before saturation. After saturation, the d$\lambda'$/d$H$ values are still negative in CFM state. The saturation field shifts to higher fields with increasing temperature, and reaches its maximum intensity at 175 K, which is close to the tricritical point in temperature scan. With further increasing temperatures, the peak gradually smears and finally disappears at 200 K, reaching the full PM state.

Based on the above determined phases and phase transition temperatures and fields in Figs. 2 and 3, an $H$-$T$ phase diagram of $Cr_2Te_3$ can be constructed, as shown in Fig. 4. The contour plots of the field-dependent d$\lambda'$/d$H$ data in Fig. 3 are served as a background for indicating the different magnetic phases. Compared with previous phase diagram obtained from the magnetization data,[10] the region to host AFM phase and the boundaries of AFM-CFM and AFM-PM are very similar while all other magnetic phases and phase boundaries are modified. At low $T$, canted FM is recognized as multi-domain canted FM and collinear FM state as the single domain canted FM. At moderate $T$, a valley of d$\lambda'$/d$H$ in the middle of AFM phase can be connected to the CFM-PM phase boundary continuously, indicating that CFM phase actually coexists with AFM phase below this boundary. From the broad feature of the valley, above this boundary, the AFM phase may coexist with a small volume of PM phase transformed from the CFM phase with short range spin-spin correlation.[12] If one can extrapolate the multiple to single domain FM phase boundary into the AFM phase, a triple point around 0.2 T and 178 K can be expected.[12] At high $T$ above 200 K, the high field region appears a clear phase boundary connected to the dome of AFM phase. A full PM phase should be located at higher temperature outside this boundary. In between PM and single-domain CFM, we expect a new magnetic phase which is dubbed as CFM2 in the phase diagram in Fig. 4.

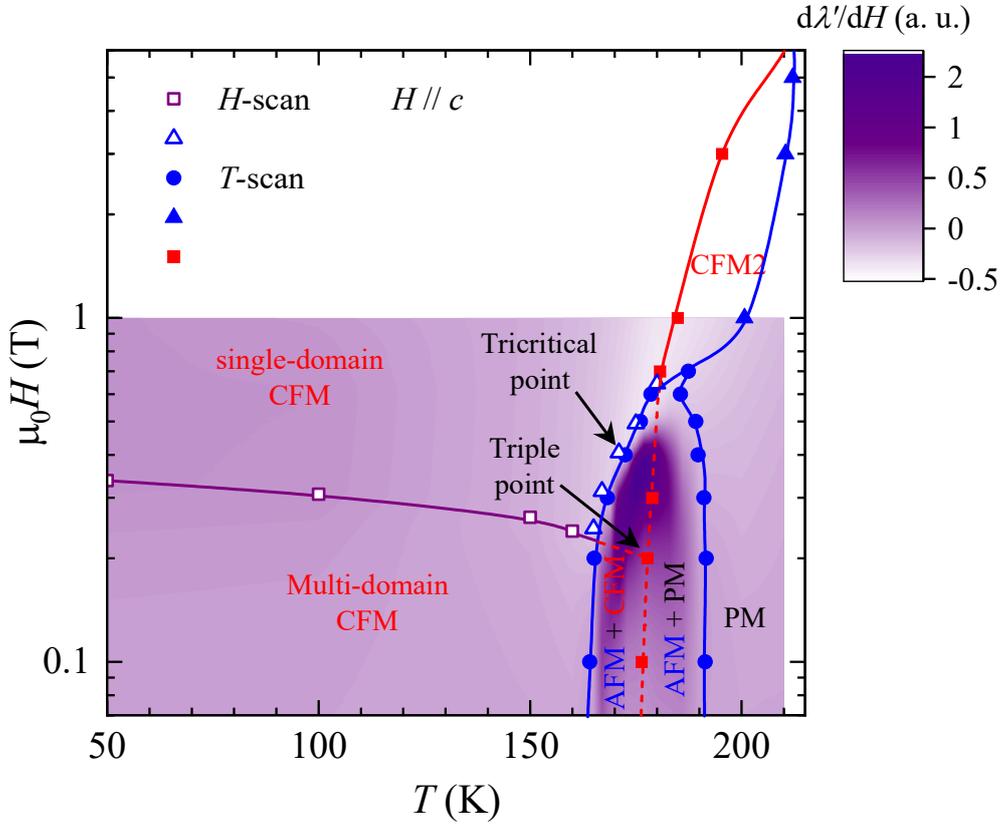

FIG. 4. An $H$-$T$ phase diagram of $Cr_2Te_3$ derived from Figs. 2 and 3. The background of the phase diagram is partially filled with the contour plots of the field-dependent $d\lambda'/dH$ data in Fig. 3. The solid and dash lines represent well-defined and proposed phase boundaries, respectively.

Finally, we should discuss the possible microscopic origin of the complicated phase coexistence and phase boundaries. In previous theoretical calculation, the ground state should be CFM state which is supported by the smaller saturation moment (2.04 $\mu_B$) than theoretical one (3.87 $\mu_B$) of $Cr^{3+}$. In the AFM state, the magnetization decreases drastically for FM-AFM transition at 165 K,[10] indicating that the Cr1/Cr3 layers must be antiparallel along the *c*-axis. In this regard, the remanent FM component inside AFM phase would be the Cr2 layer. In the higher temperature, this layer can become paramagnetic and coexist with AFM state for a small temperature region. In the high field region above 1 T, we find another magnetic phase called CFM2. It can also be the PM Cr2 layer with canted/collinear FM Cr1/Cr3 layer. Need to mention that, the high temperature transition around 200 K cannot be an impurity phase since the magnetic ordering temperatures of $CrTe_3$,[17] $CrTe_2$,[18] and $CrTe$[19] are far from this region. Therefore, our investigation reveals possibly decoupled magnetic orderings between Cr1/Cr3 layer and Cr2 layer near magnetic ordering temperature.

In conclusion, we systematically investigated the ferromagnetic $Cr_2Te_3$ single crystal system using a composite ME technique. The *H-T* phase diagram is revisited and more phases and phase boundaries are identified. CFM and AFM phases is found to overlap at high temperature regions. Meanwhile, there exist another CFM2 phase for magnetic field above 1 T. A triple point inside the AFM phase is also proposed. These findings can be attributed to the separated ordering of Cr1/Cr3 layer and Cr2 layer around 191 K and 176 K, respectively.

**ACKNOWLEDGMENTS**

This work was supported by the Natural Science Foundation of China under Grants No. 12227806, No. 11974065, No. 12274412 and No. 11874357. We would like to thank G. W. Wang and Y. Liu at Analytical and Testing Center of Chongqing University for their assistance.